\documentstyle[12pt]{article}

\makeatletter

 \@addtoreset{equation}{section}
\makeatother

\begin{document}
\topmargin -35pt
\oddsidemargin 5mm

\newcommand {\beq}{\begin{eqnarray}}
\newcommand {\eeq}{\end{eqnarray}}
\newcommand {\non}{\nonumber\\}
\newcommand {\eq}[1]{\label {eq.#1}}
\newcommand {\defeq}{\stackrel{\rm def}{=}}
\newcommand {\gto}{\stackrel{g}{\to}}
\newcommand {\hto}{\stackrel{h}{\to}}
\newcommand {\1}[1]{\frac{1}{#1}}
\newcommand {\2}[1]{\frac{i}{#1}}
\newcommand {\th}{\theta}
\newcommand {\thb}{\bar{\theta}}
\newcommand {\ps}{\psi}
\newcommand {\psb}{\bar{\psi}}
\newcommand {\ph}{\varphi}
\newcommand {\phs}[1]{\varphi^{*#1}}
\newcommand {\sig}{\sigma}
\newcommand {\sigb}{\bar{\sigma}}
\newcommand {\Ph}{\Phi}
\newcommand {\Phd}{\Phi^{\dagger}}
\newcommand {\Sig}{\Sigma}
\newcommand {\Phm}{{\mit\Phi}}
\newcommand {\eps}{\varepsilon}
\newcommand {\del}{\partial}
\newcommand {\dagg}{^{\dagger}}
\newcommand {\pri}{^{\prime}}
\newcommand {\prip}{^{\prime\prime}}
\newcommand {\pripp}{^{\prime\prime\prime}}
\newcommand {\prippp}{^{\prime\prime\prime\prime}}
\newcommand {\pripppp}{^{\prime\prime\prime\prime\prime}}
\newcommand {\delb}{\bar{\partial}}
\newcommand {\zb}{\bar{z}}
\newcommand {\mub}{\bar{\mu}}
\newcommand {\nub}{\bar{\nu}}
\newcommand {\lam}{\lambda}
\newcommand {\lamb}{\bar{\lambda}}
\newcommand {\kap}{\kappa}
\newcommand {\kapb}{\bar{\kappa}}
\newcommand {\xib}{\bar{\xi}}
\newcommand {\ep}{\epsilon}
\newcommand {\epb}{\bar{\epsilon}}
\newcommand {\Ga}{\Gamma}
\newcommand {\rhob}{\bar{\rho}}
\newcommand {\etab}{\bar{\eta}}
\newcommand {\chib}{\bar{\chi}}
\newcommand {\tht}{\tilde{\th}}
\newcommand {\zbasis}[1]{\del/\del z^{#1}}
\newcommand {\zbbasis}[1]{\del/\del \bar{z}^{#1}}
\newcommand {\vecv}{\vec{v}^{\, \prime}}
\newcommand {\vecvd}{\vec{v}^{\, \prime \dagger}}
\newcommand {\vecvs}{\vec{v}^{\, \prime *}}
\newcommand {\alpht}{\tilde{\alpha}}
\newcommand {\xipd}{\xi^{\prime\dagger}}
\newcommand {\pris}{^{\prime *}}
\newcommand {\prid}{^{\prime \dagger}}
\newcommand {\Jto}{\stackrel{J}{\to}}
\newcommand {\vprid}{v^{\prime 2}}
\newcommand {\vpriq}{v^{\prime 4}}
\newcommand {\vt}{\tilde{v}}
\newcommand {\vecvt}{\vec{\tilde{v}}}
\newcommand {\vecpht}{\vec{\tilde{\phi}}}
\newcommand {\pht}{\tilde{\phi}}
\newcommand {\goto}{\stackrel{g_0}{\to}}
\newcommand {\tr}{{\rm tr}\,}
\newcommand {\GC}{G^{\bf C}}
\newcommand {\HC}{H^{\bf C}}
\newcommand{\vs}[1]{\vspace{#1 mm}}
\newcommand{\hs}[1]{\hspace{#1 mm}}
\newcommand{\al}{\alpha}
\newcommand{\be}{\beta}
\newcommand{\Lam}{\Lambda}

\newcommand{\kahler}{K\"ahler }
\newcommand{\con}[1]{{\Gamma^{#1}}}

\setcounter{page}{0}

\begin{titlepage}

\begin{flushright}
OU-HET 407\\
PURD-TH-02-02 \\
{\tt hep-th/0203081}\\
March 2002
\end{flushright}
\bigskip

\begin{center}
{\LARGE\bf
Normal Coordinates in \kahler Manifolds

and the Background Field Method
}
\vs{10}

\setcounter{footnote}{0}
\bigskip

\bigskip
{\renewcommand{\thefootnote}{\fnsymbol{footnote}}
{\large\bf Kiyoshi Higashijima$^1$\footnote{
     E-mail: {\tt higashij@phys.sci.osaka-u.ac.jp}},
 Etsuko Itou$^1$\footnote{
     E-mail: {\tt itou@het.phys.sci.osaka-u.ac.jp}}
and Muneto Nitta$^2$\footnote{
     E-mail: {\tt nitta@physics.purdue.edu}}
}}

\vs{4}

{\sl
$^1$ Department of Physics,
Graduate School of Science, Osaka University,\\
Toyonaka, Osaka 560-0043, Japan \\

$^2$ Department of Physics, Purdue University, West Lafayette, IN
47907-1396, USA
}

\end{center}
\bigskip

\setcounter{footnote}{0}

\begin{abstract}
Riemann normal coordinates (RNC)
are unsuitable for \kahler manifolds
since they are not holomorphic.
Instead, \kahler normal coordinates (KNC) can be defined
as holomorphic coordinates.
We prove that KNC transform as a holomorphic tangent vector
under holomorphic coordinate transformations,
and therefore that they are natural extensions of RNC to the case of
\kahler manifolds.
The KNC expansion provides a manifestly covariant
background field method preserving the complex structure
in supersymmetric nonlinear sigma models.

\end{abstract}

\end{titlepage}

\section{Introduction}

The equivalence principle asserts that
general coordinate transformations on curved space-times
do not alter any physics, so that one can consider
the coordinates that make a given
application the simplest.
Riemann normal coordinates (RNC) represent one such set of coordinates
for Riemann manifolds~\cite{AFM,BB,MSV}.
They are defined as coordinates along
geodesic lines starting from a chosen point.
Hence, any point in a patch of RNC has one-to-one correspondence
with a tangent vector at the chosen point.

In most superstring theories,
extra dimensions of the higher-dimensional space-time
are compactified to a Calabi-Yau manifold~\cite{CHSW},
which is a Ricci-flat \kahler manifold.
This can be described by conformally invariant
supersymmetric nonlinear sigma models in two dimensions,
whose target spaces are \kahler manifolds~\cite{Zu}.
For perturbative (or non-perturbative) analyses,
we need to expand the Lagrangian in terms of
fluctuating fields around the background fields~\cite{perturbation}.
A generally covariant expansion
that preserves the complex structure of the target space
is most suitable in these analyses.
RNC provide a generally covariant expansion,
but they are {\it not holomorphic},
whereas \kahler normal coordinates (KNC) give us
such an expansion~\cite{HN}.
KNC are defined as coordinates satisfying
some gauge conditions on the derivatives of the metric
without recourse to geodesics~\cite{AGHKLR}.

In this paper, we prove that KNC transform
as a holomorphic tangent vector,
and therefore that they are a natural extension of RNC
to the case of \kahler manifolds.
The KNC expansion of the Lagrangian is
a manifestly covariant expansion under
holomorphic coordinate
transformations of the target space.
The relation between RNC and KNC is also shown:
we find that they differ by terms proportional to
the curvature tensor and its covariant derivatives,
and hence they coincide only in flat space.
We also give the KNC expansion of tensor fields that
can be applied to the KNC expansion of the Lagrangian
with a potential term or a higher order derivative term.

This paper is organized as follows.
In \S 2, after a short review of the RNC expansion
in Riemann manifolds, we show that RNC in \kahler manifolds
are not holomorphic and therefore are inappropriate for
an expansion preserving the holomorphy.
In \S 3, after recalling the definition of KNC,
we discuss their basic properties.
We then state a theorem elucidating
the geometrical interpretation of KNC.
The KNC expansion of tensor fields is also given.
In \S 4, we apply KNC to the background field method
in the supersymmetric nonlinear sigma models on
\kahler manifolds.
Some applications of the KNC expansion
(the Wilsonian renormalization group,
low energy theorems of Nambu-Goldstone bosons
in four dimensions, and the effective field theory on
a domain wall solution)
are discussed in \S 5.
In Appendix A, we summarize the geometry of \kahler manifolds.
The relation between RNC and KNC is discussed in detail in Appendix B.
We give a proof of the theorem presented in \S 3 in Appendix C.

\section{Riemann Normal Coordinates}
First, to compare with \kahler manifolds,
we recall some properties of RNC in Riemann manifolds,
following Ref.~\cite{AFM}.
Then, we discuss the RNC in \kahler manifolds.
It is observed that RNC are not holomorphic coordinates
in \kahler manifolds.
Therefore they are not suitable
in cases of \kahler manifolds.

\subsection{Riemann Normal Coordinates in Riemann Manifolds}
Let $\{x^A\}$ be the coordinates of a Riemann manifold $M$
($A=1,\cdots,\dim M$).
To define RNC, we choose an expansion point $\ph^A$
and consider a geodesic $\lam^A(t)$ starting from this point,
with $t$ being an affine parameter ($0 \leq t \leq 1$).
We consider the endpoint $\lam^A(1)$
as a general point $\ph^A + \pi^A$ in the manifold.
The geodesic equation in a Riemann manifold can be written
\beq
 \ddot\lam^A(t) + \con{A}_{BC}(\lam)
 \dot\lam^B(t) \dot\lam^C(t)  =  0 \, , \label{geodesics-R}
\eeq
where the dot denotes differentiation with respect to $t$,
and $\con{A}_{BC}$ is the connection.
The geodesic may be expanded in powers of the affine parameter
according to
\beq
 \lam^A (t) = \sum_{N=0}^{\infty} \1{N!} \lam^{A (N)} (0) t^N \, ,
 \label{exp-R}
\eeq
where $\lam^{A (N)}$ is the $N$-th derivative
of the geodesic with the initial condition
\beq
  \dot\lam^A(0) \equiv \xi^A \,,
\eeq
where $\xi^A$ is a tangent vector at the point $\ph^A$.
Here, $\xi^A$ is actually tangent to the geodesic.
Recursive use of the geodesic equation (\ref{geodesics-R})
gives the relations
\beq
 \lambda^{A(N)}(t)
 = -\Gamma^A{}_{B_1B_2\cdots B_N}(\lambda)
 \dot{\lambda}^{B_1}(t)\dot{\lambda}^{B_2}(t)
   \cdots\dot{\lambda}^{B_N}(t) \,.
\eeq
Here, the coefficients $\Gamma^A{}_{B_1B_2\cdots B_N}$
are defined by
\beq
\Gamma^A{}_{B_1B_2\cdots B_N}=\nabla_{B_1}\nabla_{B_2}\cdots
\nabla_{B_{N-2}}\Gamma^A{}_{B_{N-1}B_N},
\eeq
where $\nabla$ denotes the covariant derivative acting
on only lower indices of the connection.
For instance, $\con{A}_{B_1B_2B_3}$ is defined by
\beq
 \con{A}_{B_1B_2B_3} = \nabla_{B_1} \con{A}_{B_2B_3}
 = \del_{B_1} \con{A}_{B_2B_3}
 - \con{C}_{B_1B_2}\con{A}_{CB_3}
 - \con{C}_{B_1B_3}\con{A}_{B_2C} \;. \label{gamma3-R}
\eeq
We thus obtain the coefficients in (\ref{exp-R}) as
\beq
 \lam^A(t) = \ph^A + \xi^A t
 - \sum_{N=2}^{\infty}{1\over N!}
 \Gamma^A{}_{B_1B_2\cdots B_N}|_{\varphi}
 \xi^{B_1}\xi^{B_2}\cdots\xi^{B_N} t^N \, ,
 \label{lam-exp-R}
\eeq
where the index $\ph$ indicates quantities evaluated at
the initial expansion point $\ph^A$.
Since the endpoint of the geodesics is
$\ph^A + \pi^A = \lam^A(1)$, we have
\begin{equation}
 \pi^A =
 \xi^A - \sum_{N=2}^{\infty}{1\over N!}
  \Gamma^A{}_{B_1B_2\cdots B_N}|_{\varphi}
  \xi^{B_1}\xi^{B_2}\cdots\xi^{B_N} \,.
 \label{RNC-R}
\end{equation}
This can be regarded as a coordinate transformation,
and the RNC are defined by
inverting this equation to obtein $\xi^A$ as a function of $\pi$.
Therefore there is one-to-one correspondence between
a tangent vector in the tangent space at $\ph$
and a point in a patch of RNC around $\ph$.

Now, let us discuss the properties of RNC.
Since any geodesic can be written as
$\lam^A(t) = \xi^A t$ in RNC\footnote{
The degrees of freedom in
general coordinate transformations preserving $x^A = \ph^A$,
$\pi^A = x^A - \ph^A = c \pi'{}^A+\sum^{\infty}_{N=2}
c^A_{B_1\cdots B_N}\pi'{}^{B_1}\cdots \pi'{}^{B_N}$
coincide with the number of
coefficients $\Gamma^A{}_{B_1\cdots B_N}$
at (\ref{lam-exp-R}) of each order.
Hence, there exist coordinates in which
any geodesic from the origin becomes a straight line.
These coordinates are KNC.\label{straight}
},
the expansion of RNC themselves in term of the tangent vector $\xi^A$
gives the relations
\beq
 \bar \Gamma^A{}_{(B_1B_2 \cdots B_N)}|_{\ph} = 0 \,,
 \label{RNC-id.}
\eeq
where the bar indicates quantities in RNC
and the parentheses indicate the symmetrization with respect to indices:
$T_{(A_1 A_2 \cdots A_N)}=
\1{N!} (T_{A_1 A_2 \cdots A_N} + T_{A_2 A_1 \cdots A_N}
+ \cdots )$. These conditions for RNC are equivalent to
\beq
 \del_{(B_1} \del_{B_2} \cdots \del_{B_{N-2}}
 \bar \Gamma^A{}_{B_{N-1}B_N)}|_{\ph} = 0 \,. \label{RNC-id.2}
\eeq

In RNC, any tensor can be expanded in terms of
a tangent vector $\xi^A$ easily,
using the identity (\ref{RNC-id.}) or (\ref{RNC-id.2}).
For instance, the metric tensor $g_{AB}$ can be expanded as~\cite{AFM}
\beq
&& g_{AB} (x)
 = g_{AB}|_{\ph} - \1{3} R_{ACBD} |_{\ph}\xi^C \xi^D
  - \1{3!} D_{E}R_{ACBD}|_{\ph} \xi^C \xi^D \xi^E \non
&& \hs{15}
 - \1{5!} \left(6 D_C D_D R_{A E B F}
   - {16 \over 3} R_{CBD}{}^G R_{EAFG} \right)|_{\ph}
   \xi^C \xi^D \xi^E \xi^F
 + O(\xi^5)\,, \hs{10}   \label{metric-RNC}
\eeq
in which no bars appear, because
once we obtain an expansion of a tensor in RNC,
it can be regarded as an expansion in terms of a tangent vector.
Hence it is a tensor equation and holds in any coordinate system.

The RNC expansion can be applied to the background field method
of nonlinear sigma models~\cite{AFM,BB}.
(For another derivation of RNC, see Ref.~\cite{MSV}.)

\subsection{Riemann Normal Coordinates in K\"ahler Manifolds}
\label{RNC_in_Kahler}
As in Riemann manifolds considered above,
we now consider RNC in a \kahler manifold $M$.
Let $\{z^i,z^{*i}\}$ be coordinates in the K\"ahler manifold
($i=1,\cdots, \dim_{\bf C} M$).
We consider a geodesic $\lam^i(t)$ with affine parameter
$t$ ($0 \leq t \leq 1$),
starting at a point $\lam^i(0) = \ph^i$ and
ending at a point $\lam^i(1) = \ph^i + \pi^i$.
The geodesic equation in a \kahler manifold is given by
\beq
 \ddot\lam^i(t) + \con{i}_{jk}(\lam,\lam^*)
 \dot\lam^j(t) \dot\lam^k(t)  =  0 \, , \label{geodesics}
\eeq
and its complex conjugate,
where the dot denotes differentiation with respect to $t$.
In the same way, we can obtain the expansion of $\lam^i(t)$
in terms of the tangent vectors $\xi^i$ and $\xi^{*i}$.
The first few orders are given by
\beq
 && \lam^i(t) = \ph^i + \xi^i t
  - \1{2} \con{i}_{j_1j_2}|_{\ph} \xi^{j_1}\xi^{j_2} t^2
  - \1{3!} \con{i}_{j_1j_2j_3}|_{\ph}
           \xi^{j_1}\xi^{j_2}\xi^{j_3} t^3 \non
 && \hs{15}
  - \1{3!} {R^i}_{j_1k_1^*j_2}|_{\ph} \xi^{j_1}\xi^{j_2}\xi^{* k_1} t^3
  + O(t^4) \,, \label{lam-exp}
\eeq
in which $\con{i}_{j_1j_2j_3}$ is defined by
\beq
 \con{i}_{j_1j_2j_3} \equiv
 \del_{j_1} \con{i}_{j_2j_3} - \con{l}_{j_1j_2}\con{i}_{lj_3}
 - \con{l}_{j_1j_3}\con{i}_{j_2l}
 \equiv \nabla_{j_1} \con{i}_{j_2j_3} \;.  \label{gamma3}
\eeq
This is the restriction of (\ref{gamma3-R}) to holomorphic indices.
The expansion (\ref{lam-exp}) can be obtained
from the expansion (\ref{lam-exp-R})
in Riemann manifolds by identifying real coordinates
with the holomorphic and anti-holomorphic coordinates as
$\{x^A\} = \{z^i, z^{*i}\}$.

The endpoint $\ph^i + \pi^i = \lam^i(1)$
of the geodesic can be expressed by
\beq
 && \pi^i = \xi^i
  - \1{2} \con{i}_{j_1j_2}|_{\ph} \xi^{j_1}\xi^{j_2}
  - \1{3!} \con{i}_{j_1j_2j_3}|_{\ph} \xi^{j_1}\xi^{j_2}\xi^{j_3}
  - \1{3!} {R^i}_{j_1k_1^*j_2}|_{\ph} \xi^{j_1}\xi^{j_2}\xi^{* k_1} \non
 && \hs{10} + O(\xi^4) \, . \label{RNC}
\eeq
The RNC obtained by inverting this equation
depend on both $\pi$ and $\pi^*$: $\xi^i = \xi^i (\pi,\pi^*)$.
Hence, the coordinate transformation from
the holomorphic coordinates $z^i$ to the RNC $\xi^i$
is {\em not} holomorphic.
It is thus seen that
{\em Riemann normal coordinates are generally not holomorphic}.
Such non-holomorphic terms in the transformation (\ref{RNC}) appear
in conjunction with covariant tensors
like the curvature tensor ${R^i}_{j_1k_1^*j_2}$.
This is very different from the case of Riemann manifolds.

In summary, we have the following:
\begin{enumerate}

\item
The transformation (\ref{RNC}) can be directly
obtained from the transformation (\ref{RNC-R})
in Riemann manifolds with the identification of coordinates
$\{x^A\} = \{z^i, z^{*i}\}$,

\item
All non-holomorphic terms in (\ref{RNC}) appear
with coefficients of covariant tensors,
and therefore they exist in general, unless
the \kahler manifold is flat.
(See comments in \S \ref{def} and
discussion in Appendix \ref{KNC-and-RNC}.)

\end{enumerate}

\section{K\"ahler Normal Coordinates}

As shown in the last section RNC are inappropriate
for \kahler manifolds, since they are not holomorphic.
KNC~\cite{HN} are normal coordinates that are holomorphic.
In this section, after recalling the definition of KNC and giving some
discussions
in the first subsection,
we present a theorem that clarifies the
geometric properties of KNC in the second subsection.
A proof of this theorem is given in Appendix~\ref{proof}.
The KNC expansion of tensor fields is also given
in the last subsection.

\subsection{Definition of \kahler Normal Coordinates}\label{def}
Let $K(z,z^*)$ be the \kahler potential so that
$g_{ij^*}(z,z^*) = K,_{ij^*}(z,z^*)$,
where the comma denotes partial differentiation with respect
to the coordinates.
Then, decompose the coordinate $z^i$ into an expansion point $\ph^i$
and a deviation $\pi^i$ from it: $z^i = \ph^i + \pi^i$.
We define the KNC $\{\omega^i,\omega^{*i}\}$, whose origin
coincides with the expansion point $z^i = \ph^i$,
as coordinates such that the quantities
$K,_{j^*i_1\cdots i_N} = g_{i_1j^*, i_2 \cdots i_N}$
for an arbitrary $N \geq 2$
vanish at the origin of $\omega^i$ ($z^i = \ph^i$)~\cite{AGHKLR}:
\beq
 \hat K,_{j^*i_1\cdots i_N}(\omega,\omega^*)|_0 =
 \hat g_{i_1j^*, i_2 \cdots i_N}(\omega,\omega^*)|_0 = 0 \;,
 \label{KNC-cond.0}
\eeq
where the hat indicates quantities in KNC,
and the index ``$0$'' indicates a value evaluated
at the origin of KNC, $\omega^i =0$.
These conditions are equivalent to
\beq
 \del_{i_1} \cdots \del_{i_{N-2}}
  {\hat \Gamma}^j{}_{i_{N-1}i_N}
  (\omega,\omega^*)|_0 = 0 \, , \label{KNC-cond.}
\eeq
which are similar to the conditions (\ref{RNC-id.2}) for
RNC in Riemmann manifolds, except for
symmetrization with respect to indices.
The given coordinates $z^i$ (or $\pi^i$)
can be transformed to such KNC by
the {\it holomorphic} coordinate transformation~\cite{HN}
\beq
 \omega^i
 &=& \pi^i + \sum_{N=2}^{\infty} \1{N!}
 [g^{ij^*} K,_{j^*i_1 \cdots i_N}(z,z^*)]_{\ph}
   \pi^{i_1} \cdots \pi^{i_N} \non
 &=&  \sum_{N=1}^{\infty} \1{N!}
 [g^{ij^*} K,_{j^*i_1 \cdots i_N}(z,z^*)]_{\ph}
   \pi^{i_1} \cdots \pi^{i_N} \,,
 \label{KNC-def}
\eeq
where the index $\ph$ indicates that the quantity in question is evaluated
at the expansion point, $z^i =\ph^i$,
of the original coordinates.

In KNC, using (\ref{KNC-cond.0}), the K\"ahler potential can be expanded
as
\beq
&&  \hat K(\omega,\omega^*)
 = \hat K|_0 + \hat F(\omega) + \hat F^*(\omega^*)
  + \hat g_{ij^*}|_0 \omega^i \omega^{*j} \non
&& \hs{18}
 + \sum_{M,N\geq 2}
   \1{M!N!} \hat K,_{\, i_1 \cdots i_M j_1^* \cdots j_N^*}|_0
    \omega^{i_1} \cdots \omega^{i_M}
    \omega^{*j_1} \cdots \omega^{*j_N} \,, \;\;\;
 \label{KNC_exp.}
\eeq
where $\hat F(\omega)$ is a holomorphic function of $\omega$,
so that it can be eliminated by a K\"ahler transformation.
It has been shown that all coefficients of
the expansion (\ref{KNC_exp.}) are covariant tensors~\cite{HN}.
For instance, the fourth order coefficient is
$\hat K,_{i_1i_2j_1^*j_2^*}|_0 = \hat R_{i_1j_1^*i_2j_2^*}|_0$.
An explicit expression of the coefficients
in terms of the curvature tensor
and its covariant derivatives
up to sixth order is given in Ref.~\cite{HN} by
\beq
 && \hs{5}   \hat K(\omega,\omega^*) \non
 && =  \hat K|_0 + \hat F(\omega) + \hat F^*(\omega^*)
   + \hat g_{ij^*}|_0 \omega^i \omega^{*j}
   + \1{4} \hat R_{ij^*kl^*}|_0\omega^i\omega^k
                        \omega^{*j} \omega^{*l} \non
 &&+ \1{12} \hat D_m \hat R_{ij^*kl^*}|_0
    \omega^m \omega^i \omega^k \omega^{*j} \omega^{*l}
   + \1{12} \hat D_{m^*} \hat R_{ij^*kl^*}|_0
    \omega^i \omega^k \omega^{*j} \omega^{*l} \omega^{*m}\non
 && + \1{24} \hat D_n \hat D_m \hat R_{ij^*kl^*}|_0
    \omega^n\omega^m\omega^i\omega^k\omega^{*j}\omega^{*l}
   + \1{24} \hat D_{n^*} \hat D_{m^*} \hat R_{ij^*kl^*}|_0
    \omega^i\omega^k\omega^{*j}\omega^{*l}\omega^{*m}\omega^{*n}\non
 && + \1{36} \left( \hat D_{(n^*} \hat D_m \hat R_{ij^*kl^*)}
   + 3 \hat g^{or^*} \hat R_{o(j^*ml^*} \hat R_{in^*k)r^*}\right)|_0
    \omega^m\omega^i\omega^k\omega^{*j}\omega^{*l}\omega^{*n}
   + O(\omega^7)\,, \hs{8}\label{Kahler_exp}
\eeq
where $O(\omega^n)$ denotes terms of the order $n$ in
$\omega$ and $\omega^*$.
Here, the parentheses enclosing indices indicate
symmetrization with respect to the holomorphic
and anti-holomorphic indices, respectively,
e.g. $T_{(i_1i_2 \cdots i_M j_1^*j_2^* \cdots j_N^*)}
\equiv T_{(i_1i_2 \cdots i_M) (j_1^*j_2^* \cdots j_N^*)}
\equiv \1{N!M!} ( T_{(i_1i_2 \cdots i_M j_1^*j_2^* \cdots j_N^*)}
+ T_{(i_2i_1 \cdots i_M j_1^*j_2^* \cdots j_N^*)}
+ T_{(i_1i_2 \cdots i_M j_2^*j_1^* \cdots j_N^*)}
+ T_{(i_2i_1 \cdots i_M j_2^*j_1^* \cdots j_N^*)}
+ \cdots)$.\footnote{
It should be noted that we use notation that differs from that of
Ref.~\cite{HN}, in which
we used parentheses to indicate cyclic permutation
without any numerical factor.}
\footnote{
All tensors in this expansion are symmetric in
(anti-)holomorphic indices.
We do not need the symmetrization,
except for the last term,
due to the identities summarized in
Appendix \ref{kahler-mfd.}.
KNC are coordinates for which these identities become manifest.
}

~From the expansion of the \kahler potential (\ref{Kahler_exp})
we can calculate the KNC expansion of
the metric tensor through fourth order, obtaining
\beq
 &&\hs{5} \hat g_{ij^*}(\omega,\omega^*) \non
 &&= \hat g_{ij^*}|_0
   + \hat R_{ij^*kl^*}|_0\omega^k \omega^{*l}
  + \1{2} \hat D_m \hat R_{ij^*kl^*}|_0
    \omega^m \omega^k \omega^{*l}
   + \1{2} \hat D_{m^*} \hat R_{ij^*kl^*}|_0
    \omega^k \omega^{*l} \omega^{*m}\non
 && + \1{6} \hat D_n \hat D_m \hat R_{ij^*kl^*}|_0
    \omega^n\omega^m\omega^k\omega^{*l}
    + \1{6} \hat D_{n^*} \hat D_{m^*} \hat R_{ij^*kl^*}|_0
    \omega^k\omega^{*l}\omega^{*m}\omega^{*n}\non
 && + \1{4} \left(\hat D_{(n^*} \hat D_m \hat R_{ij^*kl^*)}
     + 3 \hat g^{or^*} \hat R_{o(j^*ml^*} \hat R_{in^*k)r^*}\right)|_0
    \omega^m\omega^k\omega^{*l}\omega^{*n}
    + O(\omega^5) \label{metric_exp}.
\eeq
Comparing this result with the metric expansion
in RNC (\ref{metric-RNC}),
it is seen that the coefficients in both expansions are quite different.
The relation between KNC and RNC expansions
is discussed in Appendix \ref{KNC-and-RNC}.

Let us now consider the inverse transformation of (\ref{KNC-def})
in order to compare with the transformation laws
(\ref{RNC-R}) and (\ref{RNC}) for RNC in Riemann
and \kahler manifolds, respectively.
We can show that the inverse of
the transformation (\ref{KNC-def}) is given by
\begin{equation}
 \pi^i = \omega^i - \sum_{N=2}^{\infty}{1\over N!}
 \Gamma^i{}_{j_1j_2\cdots j_N}|_{\varphi}
  \omega^{j_1}\omega^{j_2}\cdots\omega^{j_N} \;.
 \label{KNC-inverse}
\end{equation}
Here, $\Gamma^i{}_{j_1j_2\cdots j_N}$
is defined by
\beq
 \Gamma^i{}_{j_1j_2\cdots j_N}
 = \nabla_{j_1}\nabla_{j_2}\cdots
   \nabla_{j_{N-2}}\Gamma^i{}_{j_{N-1}j_N},
\eeq
in which $\nabla$ is the covariant derivative acting on the lower indices.
Equation~(\ref{KNC-inverse}) can be understood as follows:
if we take $\pi^i$ to be $\omega^i$,
we obtain the conditions
$\Gamma^i{}_{(j_1j_2\cdots j_N)}|_{\varphi}=0$,
which are actually
equivalent to the condition
(\ref{KNC-cond.0}) or (\ref{KNC-cond.}) for KNC.

We summarize this subsection as follows:
\begin{enumerate}
\item
The transformation law (\ref{KNC-inverse})
coincides with the restriction of (\ref{RNC-R})
to holomorphic indices.
In other words, only the differences between
the transformation laws (\ref{RNC}) for RNC
and (\ref{KNC-inverse}) for KNC in \kahler manifolds
consist of non-holomorphic terms associated with covariant tensors.
(See also the comments in \S \ref{RNC_in_Kahler}
and Eq.~(\ref{KNC-RNC}) in Appendix \ref{KNC-and-RNC}.)

\item
Using a holomorphic coordinate transformation,
any coordinates can be transformed into KNC,
because the freedom expressed by (\ref{KNC-inverse}) and
(\ref{KNC-def}) coincide.
However, we cannot set $\Gamma^A{}_{(B_1\cdots B_n)}|_{\ph}=0$
with any holomorphic coordinate transformation.
This is the reason that geodesics are not straight lines in KNC.
(Compare this with footnote \ref{straight} concerning the case of RNC.)

\end{enumerate}

Before closing this section, we give an example of KNC.\\
Example: A simple example of KNC is given by
the standard coordinates in the Fubini-Study metric
of ${\bf C}P^1$.
Let $z$ be a holomorphic coordinate.
Then the \kahler potential can be written as
\beq
 K(z,z^*) = \log (1 + |z|^2) \,.
\eeq
By the equation $\del_{z^*} {\del_z}^N K =
{(-1)^{N+1} N! z^{* N-1} \over (1+|z|^2)^{N+1}}$,
the condition (\ref{KNC-cond.0}) holds,
and therefore $z$ is a KNC.
Geodesics in KNC,
$z(t) = {\xi \over |\xi|} \tan( |\xi| t)
= \xi t+{1\over 3}|\xi|^2\xi t^3+\cdots$,
in which $\xi$ is a tangent vector of the geodesic,
are not linear in $t$.

\subsection{The Transformation Law of \kahler Normal Coordinates}
\label{tr-law}
RNC in a Riemann manifold
are defined by a tangent vector at the origin.
However, the geometric properties of KNC are unclear,
since KNC are not defined by geodesics.
The following theorem clarifies the geometric meaning of KNC.\\
{\bf Theorem}\\
KNC transform like a {\it holomorphic tangent vector}
at the origin of KNC, i.e.
\beq
 \omega^i \to
 \omega'{}^i
 = {\del z'{}^i \over \del z^j}\bigg|_{\ph} \omega^j \,,
 \label{theorem}
\eeq
under holomorphic coordinate transformations preserving
$z^i=\ph^i$ given by
$\pi^i \to \pi'{}^i = \pi'{}^i(\pi)
= c^i_{j_1} \pi^{j_1} + c^i_{j_1j_2}\pi^{j_1}\pi^{j_2} + \cdots$.\\
A proof of this theorem is given in Appendix \ref{proof}.

This situation is quite different from that for RNC,
because RNC transform like a tangent vector, but
they are not holomorphic in \kahler manifolds. From this theorem,
we find that
there is one-to-one correspondence in the vicinity of the origin
between a point represented by KNC and
a holomorphic tangent vector at the origin.
Therefore, KNC are a quite natural extension of RNC
to the case of a K\"ahler manifold.

We can regard the expansion (\ref{Kahler_exp}) as
an expansion in terms of a holomorphic tangent vector.
Hence (\ref{Kahler_exp}) is a {\it tensor equation} and
holds for {\it any} holomorphic coordinates $z^i$
because of the transformation law (\ref{theorem}).
The expansion of the \kahler potential around
$z^i = \ph^i$ is given by
\beq
 K(z,z^*)
 =  K|_{\ph} +  F(\omega) +  F^*(\omega^*)
   +  g_{ij^*}|_{\ph} \omega^i \omega^{*j}
   + \1{4}  R_{ij^*kl^*}|_{\ph}\omega^i\omega^k
                        \omega^{*j} \omega^{*l}
   + \cdots. \hs{2}
\eeq
Note that $z^i=\ph^i$ represents the same point in the manifold
as $\omega^i=0$ in KNC.

\subsection{The \kahler Normal Coordinate
Expansion of Tensor Fields} \label{tensor-exp.}
In this subsection we discuss the covariant expansion of
a tensor field using KNC.
Any tensor $T_{i_1 \cdots j_1^* \cdots }(z,z^*)$
can be expanded easily in KNC as
\beq
 \hat T_{i_1 \cdots j_1^* \cdots }(\omega,\omega^*)
 = \sum_{M,N=0}^{\infty} \1{M!N!}
  \hat T_{i_1 \cdots j_1^* \cdots \,},
  {}_{k_1 \cdots k_M  l_1^* \cdots l_N^*}|_0
 \omega^{k_1} \cdots \omega^{k_M}
 \omega^{*l_1} \cdots \omega^{*l_N} \,,
\eeq
where the hat indicates quantities in KNC.
All coefficients are tensors in general holomorphic coordinates.

As an example, a vector with the holomorphic index $T_i(z,z^*)$
can be expanded as
\beq
 &&\hat T_i (\omega,\omega^*)
 = \hat T_i|_0 + \hat T_{i,j}|_0 \omega^j
  + \hat T_{i,k^*}|_0 \omega^{*k}
  + \1{2} \hat T_{i,j_1j_2}|_0 \omega^{j_1} \omega^{j_2}
  + \1{2} \hat T_{i,k_1^*k_2^*}|_0 \omega^{*k_1} \omega^{*k_2} \non
 && \hs{18}
  + \hat T_{i,j_1k_1^*}|_0 \omega^{j_1} \omega^{k_1^*}
  + O(\omega^3) \,.
\eeq
Using (\ref{KNC-cond.}), each coefficient can be rewritten
as a covariant tensor in KNC as
\beq
&& \hat T_{i,j}|_0 = \hat D_{j} \hat T_i|_0 \,,\hs{5}
    \hat T_{i,k^*}|_0 = \hat D_{k^*} \hat T_i|_0 \,,\non
&& \hat T_{i,j_1j_2}|_0
     = \hat D_{j_1} \hat D_{j_2} \hat T_i|_0 \,, \hs{5}
   \hat T_{i,k_1^*k_2^*}|_0
     = \hat D_{k_1^*} \hat D_{k_2^*} \hat T_i|_0 \,, \non
 && \hat T_{i,j_1k_1^*}|_0
 = \hat D_{j_1} \hat D_{k_1^*} \hat T_i|_0
 \, (= \hat D_{k_1^*} \hat D_{j_1} \hat T_i|_0
   + \hat R^l{}_{j_1k_1^*i} \hat T_l|_0 )\,.
\eeq
(Note that covariant expressions are
not unique in general, as seen in the last equation.)
Hence the expansion of the tensor $T_i$ in terms of general holomorphic
coordinates can be obtained as
\beq
 && T_i (z,z^*)
 = T_i|_{\ph} + D_{j} T_i|_{\ph} \omega^j
  + D_{k^*} T_i|_{\ph} \omega^{*k}
  + \1{2} D_{j_1} D_{j_2} T_i|_{\ph}\omega^{j_1} \omega^{j_2} \non
&&\hs{18}
  + \1{2} D_{k_1^*} D_{k_2^*} T_i|_{\ph} \omega^{*k_1} \omega^{*k_2}
  + D_{j_1}  D_{k_1^*} T_i|_{\ph}  \omega^{j_1} \omega^{k_1^*}
  + O(\omega^3) \,,
\eeq
where no hats appear,
because this is a tensor equation, as seen from theorem (\ref{theorem}),
and therefore, it is valid in any holomorphic coordinates
(see \S \ref{tr-law}).
In the case of a holomorphic vector $T_i(z)$
[for instance $T_i(z) = \del_i W(z)$],
this expansion reduces to
\beq
 T_i (z)
 = T_i|_{\ph} + D_{j} T_i|_{\ph} \omega^j
   + \1{2} D_{j_1} D_{j_2} T_i|_{\ph}\omega^{j_1} \omega^{j_2}
   + O(\omega^3) \,. \hs{5}
\eeq
Actually, the expansion of any holomorphic tensor
$T_{i_1 \cdots i_M}(z)$ can be carried out to all orders:
\beq
 T_{i_1 \cdots i_M} (z)
 = \sum_{N=0}^{\infty} \1{N!}
  D_{j_1} \cdots D_{j_N} T_{i_1 \cdots i_M}|_{\ph}
 \omega^{j_1} \cdots \omega^{j_N} \,. \label{hol-exp.}
\eeq

In the same way, a rank two tensor $T_{ij^*}(z,z^*)$
can be expanded as
\beq
&& T_{ij^*}(z,z^*)
 = T_{ij^*}|_{\ph} + D_{k_1}T_{ij^*}|_{\ph} \omega^{k_1}
 + D_{l_1^*} T_{ij^*}|_{\ph} \omega^{*l_1} \non
&& \hs{20}
 + \1{2} D_{k_1} D_{k_2} T_{ij^*}|_{\ph} \omega^{k_1}\omega^{k_2}
 + \1{2} D_{l_1^*} D_{l_1^*} T_{ij^*}|_{\ph} \omega^{*l_1}\omega^{*l_2}
\non
&& \hs{20}
 + (D_{l_1^*} D_{k_1}  T_{ij^*} + R^m{}_{k_1l_1^*i} T_{mj^*} )|_{\ph}
    \omega^{k_1}\omega^{*l_1}
 + O(\omega^3) \,.
\eeq
In the case of the metric tensor $g_{ij^*}$, this reduces to
(\ref{metric_exp}) of this order,
because of the metric compatibility
$D_k g_{ij^*} =D_{k^*} g_{ij^*} = 0$.

\section{\kahler Normal Coordinates in
the Background Field Method}
In this section we apply the KNC expansion to
the background field method in
supersymmetric nonlinear sigma models.
The target spaces of $D=2$, ${\cal N}=2$ (or $D=4$, ${\cal N}=1$)
supersymmetric nonlinear sigma models
must be \kahler manifolds~\cite{Zu}.
The Lagrangian of supersymmetric nonlinear sigma models
with scalar fields $A^i(x)$ and Weyl fermions $\psi^i(x)$
is given (after elimination of auxiliary fields) by
(see Ref.\cite{WB})
\beq
 && {\cal L} = - g_{ij^*}(A,A^*)
  \del_{\mu} A^i \del^{\mu} A^{*j}
 - i g_{ij^*}(A, A^*)
   \psb^j \sigb^{\mu} D_{\mu} \psi^i
\non &&\hs{5}
 + \1{4} R_{ij^*kl^*} (A,A^*) \psi^i\psi^k \psb^j\psb^l\, ,
 \label{lag}
\eeq
where the covariant applied to fermions is defined by
$D_{\mu} \psi^i \equiv \del_{\mu}\psi^i
+ \del_{\mu}A^l\con{i}_{lk}(A,A^*) \psi^k$.
Scalar fields are coordinates of a \kahler manifold.
Under the holomorphic field redefinition of the scalar fields
$A^i \to A'{}^i = A'{}^i(A)$,
the fermions and the quantity $\del_{\mu} A^i(x)$
transform like holomorphic tangent vectors:
\beq
 \psi^i(x) &\to& \psi'{}^i(x)
  = {\del A'{}^i \over \del A^j} \psi^j(x) \,, \\
 \del_{\mu}A^i(x) &\to& \del_{\mu}A'{}^i(x)
  = {\del A'{}^i \over \del A^j} \del_{\mu}A^j(x) \,.
 \label{del-A}
\eeq
By its definition,
$D_{\mu} \psi^i$ also transforms like a holomorphic vector.
Therefore, the Lagrangian (\ref{lag})
is invariant under holomorphic coordinate transformations of
the target space.

Next, we consider the background field method applied to
supersymmetric nonlinear sigma models.
A manifestly supersymmetric expansion of the Lagrangian
using either RNC or KNC in terms of superfields is impossible.
If we were to promote transformation
(\ref{RNC-R}) or (\ref{KNC-def}) to
a relation between superfields,
the connection $\Gamma$ in its transformation law
would depend on both the holomorphic and anti-holomorphic coordinates
of the background, and therefore chirality
would not be preserved~\cite{Sp}.\footnote{
We take the opportunity here to correct
an error in Ref.~\cite{HN}.
A manifestly supersymmetric expansion in KNC
is impossible even around bosonic backgrounds.
A superfield expansion in KNC is possible only in
constant backgrounds.
We would like to thank Thomas E. Clark for pointing this out.
}\footnote{
In the case of a \kahler manifold with isometry,
an expansion in terms of superfields
is given by Clark and Love in Ref.~\cite{CL}
defining new holomorphic quantities.
We do not know their relation with KNC.
}
We present here the background field expansion
for the Lagrangian (\ref{lag}) in component fields
using KNC.
To this end, we decompose the complex scalar fields $A^i(x)$ into
background fields $\ph^i(x)$ and
fields $\pi^i(x)$ fluctuating around them:
\beq
 A^i(x) = \ph^i(x) + \pi^i(x) \,.
\eeq
To expand the Lagrangian in terms of the fluctuations,
we would like to transform $\pi^i(x)$ into
KNC fields $\hat \pi^i(x)$.
To do this, we must consider
the expansion of the kinetic term,
because {\it the definition of the KNC depends on
the space-time coordinates} through the background fields $\ph^i(x)$
[see Eqs.~(\ref{pi-hat}) and (\ref{deriv-exp.}), below].
This was actually recognized in
the RNC expansion in Ref.~\cite{AFM}.
Here, we generalize the treatment given in
in Ref.~\cite{AFM} to the case of KNC.

Promoting (\ref{KNC-inverse}) to a relation among fields,
the KNC fields $\hat \pi^i(x)$ can be expanded
in terms of tangent vector fields $\hat \omega^i(x)$ as
\beq
 \hat \pi^i (x) = \hat \omega^i(x)
 - \1{2} \hat\Gamma^i{}_{k_1k_2}|_{\ph}
   \hat \omega^{k_1}(x) \hat \omega^{k_2}(x)
 + O(\hat \omega^3) \, , \label{pi-hat}
\eeq
where hats indicate quantities in KNC.
When no space-time derivatives act on $\hat \pi^i$,
the KNC fields $\hat \pi^i$ coincide with
the tangent vector fields: $\hat \pi^i(x) = \hat \omega^i(x)$.
However, when the space-time derivative is applied,
the connection $\hat\Gamma^i{}_{k_1k_2}$ in (\ref{pi-hat}) is also
differentiated and remains non-zero:
\beq
 \del_{\mu} \hat \pi^i (x)
 = \hat D_{\mu} \hat \omega^i(x)
 - \1{2} \del_{\mu}\ph^{*j}(x) \hat R^i{}_{k_1j^*k_2}|_{\ph}
   \hat \omega^{k_1}(x) \hat \omega^{k_2}(x)
   + O(\hat \omega^3) \,.\label{deriv-exp.}
\eeq
We have defined the covariant derivative on a tangent vector $V^i$
at $\ph^i$ by $D_{\mu}V^i \equiv \del_{\mu}V^i
+ \del_{\mu}\ph^j \con{i}_{jk}|_{\ph}V^k$
and used the fact that it is simply $\hat D_{\mu} V^i =
\del_{\mu}V^i$ in KNC, due to (\ref{KNC-cond.}).
For general holomorphic coordinates of fluctuations $\pi^i(x)$,
Eq.~(\ref{deriv-exp.}) becomes
\beq
 \del_{\mu} \pi^i (x)
 = D_{\mu} \omega^i(x)
 - \1{2} \del_{\mu}\ph^{*j}(x) R^i{}_{k_1j^*k_2}|_{\ph}
   \omega^{k_1}(x) \omega^{k_2}(x) + O(\omega^3) \,,
 \label{deriv-exp.2}
\eeq
because of the transformation laws of (\ref{del-A})
and (\ref{theorem}).
This is a tensor equation as seen from (\ref{theorem}).

We have already given the KNC expansion of
the metric in (\ref{metric_exp}):
\beq
 g_{ij^*} (\ph + \pi, \ph^* + \pi^*)
 = g_{ij^*}|_{\ph}
 + R_{ij^*kl^*}|_{\ph} \omega^k \omega^{*l}  + O(\omega^3) \,.
 \label{metric-exp.}
\eeq
Using Eqs.~(\ref{deriv-exp.2}) and (\ref{metric-exp.}),
we obtain the expansion of the bosonic kinetic term
of the Lagrangian to
second order in the fluctuations as
\beq
 && - {\cal L}_{\rm boson}
 = g_{ij^*}|_{\ph} \del_{\mu} \ph^i \del^{\mu} \ph^{*j}
 + g_{ij^*}|_{\ph}
  ( D_{\mu} \omega^i \del^{\mu}\ph^{*j}
  + \del^{\mu}\ph^i D_{\mu} \omega^{*j})
 + g_{ij^*}|_{\ph} D_{\mu} \omega^i D_{\mu} \omega^{*j} \non
 && \hs{15}
 + R_{ij^*kl^*}|_{\ph}
  \left(\omega^k\omega^{*l} \del_{\mu} \ph^i \del^{\mu} \ph^{*j}
  - \1{2} \omega^i\omega^k \del_{\mu} \ph^{*j} \del^{\mu} \ph^{*l}
  - \1{2} \omega^{*j}\omega^{*l} \del_{\mu} \ph^i \del^{\mu} \ph^k\right)
 \non  && \hs{15} + O(\omega^3)  \,. \label{boson-exp.}
\eeq

Next, we give the expansion of the fermion kinetic term.
The expansion of the connection in KNC can be obtained from
(\ref{Kahler_exp}) as
\beq
 && \Gamma^i{}_{lk} (\ph+\pi, \ph^*+\pi^*)
 =  R^i{}_{lk_1^*k}|_{\ph} \omega^{*k_1}
  + \1{2} D_{k_2^*} R^i{}_{lk_1^*k}|_{\ph}
     \omega^{*k_1} \omega^{*k_2} \non
 && \hs{37}
 + D_{j_1} R^i{}_{lk_1^*k}|_{\ph} \omega^{j_1} \omega^{*k_1}
  + O(\omega^3) \, .
\eeq
Then, the expansion of the fermion kinetic term
to second order in $\omega$ can be obtained as
\beq
 &&- {\cal L}_{\rm fermion}
 = i g_{ij^*}|_{\ph} \psb^j \sigb^{\mu} D_{\mu} \psi^i
 + i R_{lj^*kk_1^*}|_{\ph} \del_{\mu}\ph^l
   (\psb^j \sigb^{\mu} \psi^k) \omega^{*k_1} \non
 &&\hs{18}
 + i R_{ij^*j_1k_1^*}|_{\ph}
   (\psb^j \sigb^{\mu} D_{\mu}\psi^i) \omega^{j_1} \omega^{*k_1}
 + i R_{j_1j^*kk_1^*}|_{\ph} (\psb^j \sigb^{\mu} \psi^k)
   D_{\mu} \omega^{j_1} \omega^{*k_1} \non
 &&\hs{18}
 + {i\over 2} D_{k_2^*} R_{lj^*kk_1^*}|_{\ph} \del_{\mu}\ph^l
   (\psb^j \sigb^{\mu} \psi^k) \omega^{*k_1} \omega^{*k_2} \non
 &&\hs{18}
 +  i D_{j_1} R_{lj^*kk_1^*}|_{\ph} \del_{\mu}\ph^l
   (\psb^j \sigb^{\mu} \psi^k) \omega^{j_1} \omega^{*k_2}
 + O(\omega^3) \,.  \label{fermion-exp.}
\eeq

Here we make the following comments:
\begin{enumerate}
\item
The expansions of (\ref{boson-exp.}) and (\ref{fermion-exp.})
in KNC coincide with those in RNC at this order,
since the difference between coordinates first appears
at third order, as seen in Eq.~(\ref{rel}).
To preserve holomorphic structures beyond this order,
we must use the KNC expansion given in this section.

\item
In a constant background, with $\del_{\mu}\ph^i = 0$,
the expansion of (\ref{boson-exp.}) and (\ref{fermion-exp.})
reduces to the expansion given in Ref.~\cite{HN}.

\end{enumerate}

Supersymmetric nonlinear sigma models possess the
potential term~\cite{WB}
\beq
&& {\cal L}_{\rm potential}
 = - g^{ij^*}(A,A^*) D_i W(A) D_{j^*} W(A^*) \non
&&\hs{20}
   - \1{2} D_iD_j W(A) \psi^i \psi^j
   - \1{2} D_{i^*}D_{j^*} W^*(A^*) \psb^i \psb^j \,,
\eeq
where $W(A)$ is a holomorphic function called a ``superpotential''.
Using (\ref{hol-exp.})
and the relation $g^{ij^*}(A,A^*) = g^{ij^*}|_{\ph}
+ R^{ij^*}{}_{kl^*}|_{\ph} \omega^k\omega^{*l} + O(\omega^3)$,
the potential term can be expanded as
\beq
&& - {\cal L}_{\rm potential}
 = [g^{ij^*} D_i W D_{j^*} W^* ]_{\ph} \non
&& + [g^{ij^*} (D_{k_1} D_i W) D_{j^*} W^* ]_{\ph} \omega^{k_1}
 + [g^{ij^*} D_i W (D_{l_1^*} D_{j^*} W^*) ]_{\ph} \omega^{*l_1} \non
&&
 + \1{2} [g^{ij^*} (D_{k_1}D_{k_2} D_i W) D_{j^*} W^*]_{\ph}
     \omega^{k_1}\omega^{k_2}
 +  \1{2} [g^{ij^*} D_i W (D_{l_1^*}D_{l_2^*}D_{j^*} W^*)]_{\ph}
     \omega^{*l_1}\omega^{*l_2} \non
&& + [g^{ij^*}  (D_{k_1} D_i W)(D_{l_1^*} D_{j^*} W^*)
   + R^{ij^*}{}_{k_1l_1^*}  D_i W D_{j^*} W^*]|_{\ph}
   \omega^{k_1} \omega^{*l_1}  \non
&& + \1{2} (D_iD_j W|_{\ph} + D_{k_1}D_iD_jW|_{\ph} \omega^{k_1}
    + \1{2} D_{k_1}D_{k_2}D_iD_jW|_{\ph}\omega^{k_1}\omega^{k_2})
    \psi^i \psi^j \non
&& + \1{2} (D_{i^*}D_{j^*} W^*|_{\ph}
          + D_{k_1^*}D_{i^*}D_{j^*}W^*|_{\ph} \omega^{*l_1}
    + \1{2} D_{k_1^*}D_{k_2^*}D_{i^*}D_{j^*}W^*|_{\ph}
          \omega^{*l_1}\omega^{*l_2})
        \psb^i \psb^j  \non
&& + O(\omega^3) \,. \label{pot.-exp.}
\eeq

\section{Discussion}
We discuss some applications of
the KNC expansion in this section.

1. Nonlinear sigma models are renormalizable in two dimensions.
Perturbative methods~\cite{perturbation},
however, cannot be used in the large coupling regime.
On the contrary, the Wilsonian renormalization
group (WRG)~\cite{WRG}
can be applied in this region, and it may lead new results.
To derive the WRG equation, we need to expand the Lagrangian around
the background field.
Using the KNC expansion, we can derive a WRG equation
that is generally covariant under the reparameterization of
the background field.
This would provide a better understanding of
non-perturbative aspects of
supersymmetric nonlinear sigma models,
combined with non-perturbative analysis of
Hermitian symmetric spaces
using the large $N$ method~\cite{large-N}
and related models applied to Ricci-flat \kahler manifolds~\cite{HKN}.

2. In four dimensions, nonlinear sigma models can be
considered as effective field theories
corresponding to theories at higher energy scales,
such as supersymmetric QCD or
the minimal supersymmetric standard model.
When symmetry is spontaneously broken down,
there appear massless (quasi-)Nambu-Goldstone bosons in addition to
fermionic superpartners~\cite{SSB}.
Using the KNC expansion,
low energy theorems of scattering amplitudes
for these bosons are studied in Ref.~\cite{HNOO}.
A manifestly supersymmetric
four derivative term with a rank four tensor was recently
reported in Ref.~\cite{Ni}.
Hence it appears possible to obtein a supersymmetric extension of
the chiral perturbation theory
by applying the KNC expansion of tensor fields
given in \S \ref{tensor-exp.} to
higher rank tensors.

3. Some supersymmetric nonlinear sigma models
with suitable potentials
admit BPS domain wall solutions,
which break a half of the original supersymmetry
(see, e.g., Ref.~\cite{NNS}).
The effective field theory on a wall is very interesting
in the brane world scenario.
To obtain this,
we need to expand the Lagrangian around
the domain wall background,
as was done in the case of
linear models in Ref.~\cite{CS}.
We believe that the KNC expansion
[with the potential term (\ref{pot.-exp.})]
will be found to be a very powerful tool
to construct effective field theories on BPS domain walls in
supersymmetric nonlinear sigma models.

\section*{Acknowledgements}
We are grateful to Thomas E. Clark for useful discussions
and for reading the manuscript.
We also would like to thank Hisao Suzuki for a valuable suggestion,
which was one of the motivations of this work.
M.~N. thanks Masato Arai, Masashi Naganuma and Norisuke Sakai
for discussions on the application of KNC to
the BPS domain walls.
This work was supported in part by
a Grant-in-Aid for Scientific Research.
The work of M.~N. was supported by the U.S.
Department of Energy under grant DE-FG02-91ER40681 (Task B).

\begin{appendix}
\section{\kahler Manifolds}\label{kahler-mfd.}
In this appendix,
we summarize the geometry of K\"{a}hler manifolds.
Here, uppercase Roman letters are used for
both holomorphic and anti-holomorphic indices:
$\{x^A\} = \{z^i, z^{*i}\}$.
The the complex structure $J$ and
the Hermitian metric $g$ are given
on Hermitian manifolds.
The \kahler form $\Omega \equiv i g_{ij^*} dz^i \wedge dz^{*j}$
is closed on \kahler manifolds: $d \Omega =0$. From this condition,
the metric can be written as
\beq
 g_{ij^*}(z,z^*)
  = {\del^2 K(z,z^*) \over \del z^i \del z^{*j}}
  = K,_{ij^*} (z,z^*),
\eeq
using a real function $K$ called the \kahler potential.
There exists an ambiguity in the definition of $K$
in the sense that given \kahler potential $K$,
$K' = K + f(z) + f^*(z^*)$, with arbitrary holomorphic function $f$,
is also a \kahler potential.

Components of the affine connection with mixed indices disappear
as a result of the compatibility condition
of the complex structure, $D J = 0$.
The non-zero components of the connection are
\beq
 {\Gamma^k}_{ij} = g^{kl^*} g_{jl^*, i}
                 = g^{kl^*} K,_{\,ijl^*}
\eeq
and their conjugates.
The independent components of the curvature tensor are
\beq
 {R^{i^*}}_{j^*kl^*}
  = \del_k \Gamma^{i^*}{}_{j^*l^*}
  = \del_{k}(g^{mi^*} g_{mj^*,l^*})
\eeq
and thier conjugates.
The curvature tensor with lower indices
\beq
 R_{ij^*kl^*}
 \equiv g_{im^*} {R^{m^*}}_{j^*kl^*}
 =  K,_{\,ij^*kl^*}
 - g^{mn^*} K,_{\,mj^*l^*} K,_{\,n^*ik}
\eeq
has some symmetries among its indices.
In addition to the symmetries of the curvature tensor
on Riemann manifolds,
\beq
 R_{ABCD} = - R_{ABDC} = - R_{BACD} = R_{CDAB} \,,
\eeq
there exist the symmetries
\beq
 R_{ij^*kl^*} = R_{kj^*il^*} = R_{il^*kj^*},
\eeq
as a result of the \kahler condition.

The Bianchi identity
$D_{A} R_{BCDE}+ D_{C} R_{ABDE} + D_{B} R_{CADE} = 0$
on Riemann manifolds reduces to
\beq
 D_m R_{ij^*kl^*} = D_i R_{mj^*kl^*} \,
\eeq
in the case of \kahler manifolds.
Commutators of covariant derivatives of an arbitrary tensor
$T_{C_1 \cdots  C_n}$  are given by
\beq
 [D_A,D_B] \,T_{C_1 \cdots C_n}
= \sum_{a=1}^n
  {R_{ABC_a}}^D T_{C_1 \cdots C_{a-1} D C_{a+1} \cdots C_n} \,.
   \label{com.D-D}
\eeq
The equations $[D_i,D_j] = [D_{i^*},D_{j^*}] =0$ hold
as a result of the K\"{a}hler property.
KNC are the coordinates such that these identities
become manifest.

\section{Relation Between \kahler and
Riemann Normal Coordinates}\label{KNC-and-RNC}
In \S \ref{RNC_in_Kahler}, we considered geodesics
in general holomorphic coordinates.
Considering geodesics in the KNC $\omega^i$
instead of the general coordinates $\pi^i = z^i - \ph^i$,
we can obtain the relation between KNC and RNC.
Their relation up to fourth order is obtained
instead of (\ref{RNC}) as
\beq
 && \omega^i = \xi^i
 - \1{3!} \hat R^i{}_{j_1k_1^*j_2}|_0 \xi^{j_1}\xi^{j_2}\xi^{* k_1}
 - {2\over 4!} \hat D_{j_1}\hat R^i{}_{j_2k_1^*j_3}|_0
    \xi^{j_1}\xi^{j_2}\xi^{j_3}\xi^{* k_1} \non
 && \hs{10} - \1{4!} \hat D_{k_1^*}\hat R^i{}_{j_1k_2^*j_2}|_0
    \xi^{j_1}\xi^{j_2}\xi^{* k_1}\xi^{* k_2} + O(\xi^5) \,,
 \label{rel}
\eeq
where the hat indicates quantities in KNC,
and the condition (\ref{KNC-cond.}) has been used.
In general, all coefficients in the expansion of
the transformation from
RNC to KNC are covariant tensors $T$ composed of
the curvature and metric tensors and
their covariant derivatives, as follows from (\ref{KNC-cond.}):
\beq
 \omega^i = \xi^i - \sum_{M =2, N = 1}^{\infty}
  \hat T^i{}_{j_1 \cdots j_M k_1^* \cdots k_N^*}
  (\hat D, \hat R, \hat g)|_0
  \xi^{j_1} \cdots \xi^{j_M}
  \xi^{*k_1} \cdots \xi^{*k_N} \,. \label{KNC-RNC}
\eeq

Here we give some comments:
\begin{enumerate}
\item
KNC and RNC coincide if and only if the \kahler manifold is flat.

\item
In the case of Hermitian symmetric spaces,
the equation $DR=0$ holds.
Therefore the tensors $T$ in (\ref{KNC-RNC}) are composed
of only the curvature and metric tensors.
\end{enumerate}

Next we demonstrate the relation between KNC and RNC with regard to
the expansion of the metric tensor
in these coordinates. From the transformation law (\ref{rel}),
the Jacobian can be calculated to give
\beq
 && {\del \omega^i \over \del \xi^l}
 = \delta^i_l
 - \1{3} \hat R^i{}_{j_1k_1^*l}|_0\xi^{j_1}\xi^{* k_1}
 - {1\over 4} \hat D_{j_1}\hat R^i{}_{j_2k_1^*l}|_0
    \xi^{j_1}\xi^{j_2} \xi^{* k_1} \non
 &&\hs{15}
 - \1{12} \hat D_{k_1^*}\hat R^i{}_{j_1k_2^*l}|_0
    \xi^{j_1} \xi^{* k_1}\xi^{* k_2} + O(\xi^4) \,, \non
 && {\del \omega^i \over \del \xi^{*l} }
 =
 - \1{6} \hat R^i{}_{j_1l^*j_2}|_0\xi^{j_1}\xi^{j_2}
 - {1\over 12} \hat D_{j_1}\hat R^i{}_{j_2l^*j_3}|_0
    \xi^{j_1}\xi^{j_2}\xi^{j_3} \non
 &&\hs{15}
 - \1{12} \hat D_{k_1^*}\hat R^i{}_{j_1l^*j_2}|_0
    \xi^{j_1}\xi^{j_2}\xi^{* k_1} + O(\xi^4) \,.
 \label{Jacobian}
\eeq
The KNC expansion of the metric (\ref{metric_exp}) is transformed to
\beq
&& \bar g_{ij^*}
 = \hat g_{ij^*}|_0 - \1{3} \hat R_{ij^*kl^*}|_0 \xi^k \xi^{*l}
 - \1{6} \hat D_{m} \hat R_{ij^*kl^*}|_0\xi^m\xi^k\xi^{*l}
 \non && \hs{15}
 - \1{6} \hat D_{m^*} \hat R_{ij^*kl^*}|_0\xi^k\xi^{*l}\xi^{*m}
 + O(\xi^4)\, \non
&& \bar g_{ij}
 =  - \1{3} \hat R_{ik^*jl^*}|_0 \xi^{*k} \xi^{*l}
 - \1{6} \hat D_{k^*} \hat R_{il^*jm^*}|_0\xi^{*k}\xi^{*l}\xi^{*m}
\non && \hs{15}
 - \1{6} \hat D_{k} \hat R_{il^*jm^*}|_0\xi^k\xi^{*l}\xi^{*m}
 + O(\xi^4)\,  , \label{RNC-metric}
\eeq
where the bar indicates the tensors in RNC.
Note that the tensors on the right-hand sides of these equations
are tensors in KNC (or general holomorphic coordinates).
There appear non-Hermitian components,
$\bar g_{ij}$ and its conjugate,
since the transformation (\ref{KNC-RNC}) is not holomorphic.
The components given in (\ref{RNC-metric}) coincide
with the RNC expansion of the metric (\ref{metric-RNC}),
identifying real coordinates as $\{x^A\}=\{z^i,z^{*i}\}$ and
enforcing the \kahler condition on the curvature tensor
as $R_{ijkl}=R_{ij^*kl}=R_{ijkl^*}=0$
on the right-hand side of (\ref{metric-RNC}).
We would like to emphasize again that
the RNC expansion of the metric
includes unwanted non-Hermitian terms.

\section{A Proof of the Theorem}\label{proof}

In this appendix,
we show that KNC in a K\"ahler manifold
can be interpreted as a holomorphic tangent vector
at the origin,
and therefore they are a natural extension of
RNC to the case of a K\"ahler manifold.
To this end, we consider the relation between
a set of KNC defined by a set of
general holomorphic coordinates
$z^i$ and $z'{}^i = z'{}^i(z)$,
which are transformed under
a holomorphic coordinate transformation preserving the origin.
(In this appendix, we take the expansion point to be the origin,
$\ph^i=0$, for simplicity,
but the entire treatment holds for general expansion points,
replacing $z^i$ by $\pi^i = z^i - \ph^i$.)

First, we need the transformation law of
the ``generalized connection''
$K,_{j^*i_1 \cdots i_N} (z,z^*)$ in
the definition of KNC, given by the following lemma.\\
{\bf Lemma}\\
The transformation law of
$K,_{j^*i_1 \cdots i_N} (z,z^*)$
under a holomorphic coordinate transformation
$z^i \to z'{}^i = z'{}^i(z)$ is given by
\beq
 \hs{-5}
 &&K,_{j^*i_1 \cdots i_N}(z,z^*) \non
 &\to&
 K,_{j'{}^* i_1' \cdots i_N'}(z',z'{}^*)
 = \sum_{n=1}^N \1{n !} K,_{l^*k_1 \cdots k_n} (z,z^*)
 {\del z^{*l} \over \del z'{}^{*j}}
 \left[ {\del^N (z^{k_1} \cdots z^{k_n})
  \over \del z'{}^{i_1} \cdots \del z'{}^{i_N} } \right]_* ,
 \;\;\;\;\;
 \label{lemma}
\eeq
where $[\cdots]_*$ possesses the meaning that
terms including $z$ that are differentiated by no $z'$
are omitted.

The term for $n=N$ is a homogeneous (tensorial) term,
but all of the other terms are non-homogeneous terms.
The $N=2$ case corresponds to the ordinary connection:
$K,_{j^*i_1i_2} = g_{kj^*}\Gamma^k_{i_1i_2}$.
\\
(Proof) We use mathematical induction for the proof.\\
i) First, we consider the $N=1$ case.
In this case, Eq.~(\ref{lemma}) is
\beq
 K,_{j^*i_1} \to
 K,_{j'{}^* i_1'} = K,_{l^*k_1}
 {\del z^{*l} \over \del z'{}^{*j}}
 {\del z^{k_1} \over \del z'{}^{i_1}} \,.
\eeq
This is obvious, because $K,_{j^*i_1} = g_{i_1 j^*}$.\\
ii) We assume that Eq.~(\ref{lemma}) holds for $N$.
Differentiation of Eq.~(\ref{lemma})
with respect to $z'{}^{i_{N+1}}$ gives
\beq
 K,_{j'{}^* i_1'\cdots i_{N+1}'} &=&
 \sum_{n=1}^N
 \1{n !} K,_{l^*k_1 \cdots k_{n+1}}
 {\del z^{*l} \over \del z'{}^{*j}}
 {\del z^{k_{n+1}} \over \del z'{}^{i_{N+1}}}
 \left[ {\del^N (z^{k_1} \cdots z^{k_n})
  \over \del z'{}^{i_1} \cdots \del z'{}^{i_N} } \right]_* \non
&& + \sum_{n=1}^N
 \1{n !} K,_{l^*k_1 \cdots k_n}
 {\del z^{*l} \over \del z'{}^{*j}}
 {\del \over \del z'{}^{i_{N+1}}}
 \left[ {\del^N (z^{k_1} \cdots z^{k_n})
  \over \del z'{}^{i_1} \cdots \del z'{}^{i_N} } \right]_*\,.
\eeq
The first term can be rewritten as
\beq
 \sum_{n=2}^{N+1} {1 \over (n-1)!}
 K,_{l^*k_1 \cdots k_n}
 {\del z^{*l} \over \del z'{}^{*j}}
 {\del z^{k_n} \over \del z'{}^{i_{N+1}}}
 \left[ {\del^N (z^{k_1} \cdots z^{k_{n-1}})
  \over \del z'{}^{i_1} \cdots \del z'{}^{i_N} } \right]_* \,.
\eeq
Therefore, we have
\beq
 K,_{j'{}^* i_1'\cdots i_{N+1}'}
&=& K,_{l^*k_1} {\del z^{*l} \over \del z'{}^{*j}}
 {\del^{N+1} z^{k_1}
  \over \del z'{}^{i_1} \cdots \del z'{}^{i_{N+1}} } \non
 &&+ \sum_{n=2}^N {1 \over n!}
 K,_{l^*k_1 \cdots k_n}
 {\del z^{*l} \over \del z'{}^{*j}}
 \left\{
 n {\del z^{k_n} \over \del z'{}^{i_{N+1}}}
 \left[ {\del^N (z^{k_1} \cdots z^{k_{n-1}})
  \over \del z'{}^{i_1} \cdots \del z'{}^{i_N} } \right]_*
 \right. \non
 && \left.\hs{42}
 + {\del \over \del z'{}^{i_{N+1}}}
 \left[ {\del^N (z^{k_1} \cdots z^{k_n})
  \over \del z'{}^{i_1} \cdots \del z'{}^{i_N} } \right]_*
 \right\} \non
&&+ \1{N !} K,_{l^*k_1 \cdots k_{N+1}}
 {\del z^{*l} \over \del z'{}^{*j}}
 {\del z^{k_{N+1}} \over \del z'{}^{i_{N+1}}}
 \left[ {\del^N (z^{k_1} \cdots z^{k_N})
  \over \del z'{}^{i_1} \cdots \del z'{}^{i_N} } \right]_* \,.
\eeq
With regard to the term in the curly brackets on the right-hand side, the
relation
\beq
 && n {\del z^{k_n} \over \del z'{}^{i_{N+1}}}
 \left[ {\del^N (z^{k_1} \cdots z^{k_{n-1}})
  \over \del z'{}^{i_1} \cdots \del z'{}^{i_N} } \right]_*
 + {\del \over \del z'{}^{i_{N+1}}}
 \left[ {\del^N (z^{k_1} \cdots z^{k_n})
  \over \del z'{}^{i_1} \cdots \del z'{}^{i_N} } \right]_* \non
 && = \left[ {\del^{N+1} (z^{k_1} \cdots z^{k_n})
  \over \del z'{}^{i_1} \cdots \del z'{}^{i_{N+1}} } \right]_* \
\eeq
holds, where symmetrization of the first term
on the left-hand side is implied.
We thus obtain
\beq
 K,_{j'{}^* i_1'\cdots i_{N+1}'}
 = \sum_{n=1}^{N+1} \1{n !} K,_{l^*k_1 \cdots k_n}
 {\del z^{*l} \over \del z'{}^{*j}}
 \left[ {\del^{N+1} (z^{k_1} \cdots z^{k_n})
  \over \del z'{}^{i_1} \cdots \del z'{}^{i_{N+1}} } \right]_* \,.
\eeq
iii) From i) and ii) the lemma is proved. (Q.E.D.)

We call a holomorphic coordinate transformation that
leaves the origin invariant (i.e. $z^i=0$ implies $z'{}^i=0$
and vice versa),
a ``holomorphic coordinate transformation preserving the origin''.
We immediately obtain the following corollary from the lemma:\\
{\bf Corollary}\\
Under holomorphic coordinate transformations preserving the origin,
$K,_{j^*i_1 \cdots i_N}$ transforms according to
\beq
 K,_{j^*i_1 \cdots i_N}|_0 \to
 K,_{j'{}^* i_1' \cdots i_N'}|_0
 &=& \sum_{n=1}^N \1{n !} K,_{l^*k_1 \cdots k_n}|_0
 \left[
 {\del z^{*l} \over \del z'{}^{*j}}
 {\del^N (z^{k_1} \cdots z^{k_n})
  \over \del z'{}^{i_1} \cdots \del z'{}^{i_N} } \right]_0  \non
 &=& \sum_{n=1}^{\infty} \1{n !} K,_{l^*k_1 \cdots k_n}|_0
 \left[
 {\del z^{*l} \over \del z'{}^{*j}}
 {\del^N (z^{k_1} \cdots z^{k_n})
  \over \del z'{}^{i_1} \cdots \del z'{}^{i_N} } \right]_0
 \label{corollary}
\eeq
at the origin,
where the subscripts ``$0$'' indicate that the values are evaluated at
the origin: $z^i=0$ or $z'{}^i=0$.
The second equality holds because
the term $[\cdots]_0$ vanishes when $n>N$.

We are now ready to prove the theorem,
which reveals the geometric meaning of KNC.\\
{\bf A proof of the theorem:}
Using the definition (\ref{KNC-def}) and
the corollary (\ref{corollary}),
the left-hand side of Eq.~(\ref{theorem}) can be explicitly
calculated as
\beq
 \omega'{}^i
&=& \sum_{n=1}^{\infty} \1{n!}
 (g^{\prime ij^*} K,_{j'{}^*i_1' \cdots i_n'})_0
   z'{}^{i_1} \cdots z'{}^{i_n} \non
&=& \sum_{n=1}^{\infty} \1{n!} g^{\prime ij^*}|_0
 \left(\sum_{m=1}^{\infty} \1{m !} K,_{l^*k_1 \cdots k_m}|_0
 \left[
 {\del z^{*l} \over \del z'{}^{*j}}
 {\del^n (z^{k_1} \cdots z^{k_m})
  \over \del z'{}^{i_1} \cdots \del z'{}^{i_n} } \right]_0\right)
 z'{}^{i_1} \cdots z'{}^{i_n} \non
&=& {\del z'{}^i \over \del z^k}\bigg|_0
 \sum_{n=1}^{\infty} \sum_{m=1}^{\infty} \1{n!m!}
 (g^{kl^*} K,_{l^*k_1 \cdots k_m})_0
 \left[{\del^n (z^{k_1} \cdots z^{k_m})
  \over \del z'{}^{i_1} \cdots \del z'{}^{i_n} } \right]_0
 z'{}^{i_1} \cdots z'{}^{i_n}  \non
&=& {\del z'{}^i \over \del z^k}\bigg|_0
 \sum_{m=1}^{\infty} \1{m!}
 (g^{kj^*} K,_{j^*k_1 \cdots k_m})_0
 z^{k_1} \cdots z^{k_m}
= {\del z'{}^i \over \del z^k}\bigg|_0  \omega^k \,.
\eeq
(Q.E.D.)

\end{appendix}


\end{document}